\documentstyle[epsf,psfig,natbib_mod]{mn}
\begin{document}
\title[NGC~4051 in the low flux state]
{Complex X-ray spectral behaviour of NGC~4051 in the low flux state}
\author[P. Uttley et al.]
  {P. Uttley$^1$\thanks{E-mail: pu@astro.soton.ac.uk},
R. D. Taylor$^1$, I. M.~M$^{\rm c}$Hardy$^1$, M. J. Page$^2$,
K. O. Mason$^2$,\newauthor G. Lamer$^3$ and A. Fruscione$^4$ \\
  $^1$School of Physics and Astronomy, University of Southampton,
Southampton SO17 1BJ \\
  $^2$Mullard Space Science Laboratory, University College London,
Holmbury St Mary, Dorking RH5 6NT \\
$^3$Astrophysikalisches Institut Potsdam, An der
Sternwarte 16, D-14482 Potsdam, Germany\\
$^4$Harvard-Smithsonian Center for Astrophysics, 60
Garden Street, Cambridge, MA 02138, USA\\
}
\date{Accepted 2003 October 23.  Received 2003 September 23}
\maketitle
\parindent 18pt

\begin{abstract}
The Narrow Line Seyfert~1 galaxy NGC~4051 was observed in one of its
prolonged low-lux states by {\it XMM-Newton} in November 2002.  Here
we present the results of an analysis of EPIC-pn data obtained
during the observation.  Within the low state, the source shows
complex spectral variability which cannot easily be explained by any
simple model.  However, by making a `flux-flux' plot which combines
the low state data with data obtained during a normal
flux state, we demonstrate that the extremely hard spectrum observed above
2~keV results from a continuation of the spectral variability seen in
the normal state, which is caused by spectral pivoting of the
power-law continuum.  The pivoting power-law appears to be attached to
a Comptonised thermal component of variable flux (blackbody temperature
$kT\sim0.1$~keV, consistent with the small black
hole mass in NGC~4051) which dominates the soft X-ray
band in the low state, and is probably the source of seed photons for
Comptonisation.  Additional constant thermal and reflection components,
together with absorption by
ionised gas, seem to be required to complete the picture and explain
the complex X-ray spectral variability seen in the low state of
NGC~4051.
\end{abstract}


\section{Introduction}
NGC~4051 is a low luminosity (typically few $10^{41}$~erg~s$^{-1}$) 
Narrow Line Seyfert~1 (NLS~1) AGN which shows extreme X-ray flux
variability and associated strong spectral variability
(e.g. \citealt{gua96},\citealt{lam03a}), on both long and short
time-scales.  In particular, the source shows unusual low flux states,
lasting weeks to months, during which the X-ray spectrum becomes
extremely hard (photon index $\Gamma\sim1$) above a few keV 
but is dominated by a much softer component ($\Gamma\sim3$)
at lower energies (\citealt{gua98}, \citealt{utt99}).  Previously, we
reported results of a {\it Chandra} CCD
observation of NGC~4051 in the low flux state (\citealt{utt03},
henceforth U03) which revealed that, even in the low state, the hard and soft components were
significantly variable and correlated with one another, and so could not originate
primarily in extended emission, such as reflection from a torus and extended
scattering medium.  The {\it Chandra} image also clearly ruled out any significant
extended emission on $\sim100$~pc scales (confirming an earlier result
based on a {\it Chandra} grating observation obtained in a normal flux
state, \citealt{col01}).  In U03 we further noted
that the unusual curvature at harder energies in the 
{\it Chandra} spectrum, when compared with a spectrum of the low state
obtained by the {\it Rossi X-ray Timing Explorer} ({\it RXTE}), was
consistent with the presence of a very prominent gravitationally
redshifted diskline, suggesting that the reflection features from
close to a black hole may remain constant in flux in NGC~4051 despite
large changes in the continuum flux, as appears also to
be the case in MCG--6-30-15 (\citealt{fab03},\citealt{tay03}).

In U03 we suggested that the unusual
spectral shape of the low state was simply a continuation to low
fluxes of the normal spectral variability,
i.e. with the spectrum hardening towards lower fluxes.
 In other words, the low flux state is probably not a physically distinct
state, in the sense used to describe the states of e.g. X-ray
binaries, and the extreme long-term X-ray spectral 
variability of NGC~4051 is likely produced by the same
physical process as the short-term spectral variability.  Using a
model-independent method, the `flux-flux' plot,
\citet{tay03} find that the spectral variability of NGC~4051 in
the 2-15~keV band is best
explained by pivoting of the power-law continuum about an energy of
$\sim100$~keV, in addition to a hard constant component.  This result
contrasts with the X-ray spectral variability of other Seyferts, which
is best explained by a constant hard component together with a
variable component which does not pivot but
maintains a constant shape (\citealt{tay03}, \citealt{fab03}).

Observations of NGC~4051 in the low
state allow us to test the limits of the spectral pivoting model
suggested by \citet{tay03}.  In this paper, we present the results of
an {\it XMM-Newton} Target of Opportunity Observation (TOO) 
of NGC~4051 obtained towards the
end of a recent low state.  We first demonstrate the complex
spectral variability in the low state and then combine the low state
data with data obtained during a
normal state, to make a `flux-flux' plot \citep{tay03} to
test the pivoting model and determine whether the low state spectrum is
consistent with spectral pivoting down to the lowest observed fluxes.
We also investigate the presence of reflection features in the hard
spectrum obtained during the low state, and then 
combine these results with the inferences we make
from the flux-flux plots, to construct a simple broadband spectral
model which can explain the observed spectral shape and variability of
NGC~4051.  Using spectral fits, we demonstrate how our model can explain the complex
spectral variability observed within the low state.  
We will present spectral fits of our model over the entire observed flux 
range of NGC~4051 (i.e. including the normal state) in a future paper
(Taylor et al., in prep.).

\begin{figure}
\begin{center}
{\epsfxsize 0.9\hsize
 \leavevmode
 \epsffile{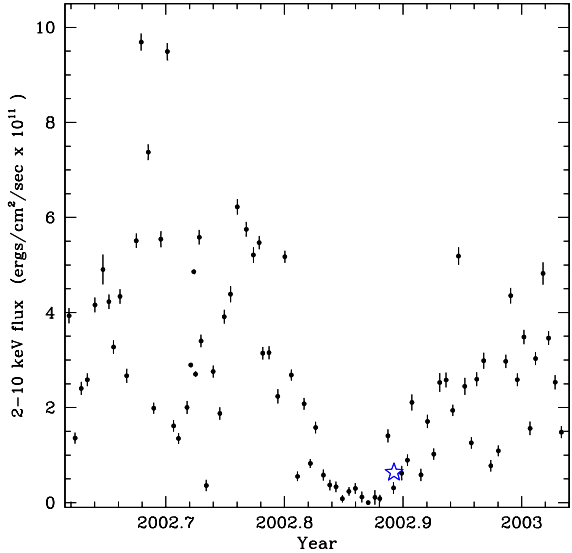}
}\caption{Recent 2-10~keV light curve of NGC~4051, obtained with {\it
RXTE} monitoring (see \citealt{lam03a}, \citealt{mch03} for details).
The time of the {\it XMM-Newton} TOO observation (and the
2-10 keV flux observed by {\it XMM-Newton}) is marked by a star.}
\label{longlc}
\end{center}
\end{figure}

\section{Observation}
\label{obs}
Our continuing {\it RXTE} monitoring observations showed that NGC~4051 entered a low state,
characterised by low flux and a small variability amplitude, for about
a month in 2002 October-November (Fig.~\ref{longlc}).   On detecting
that the source was in a low state with {\it RXTE}, we triggered a TOO observation with {\it
XMM-Newton}, which observed the
source on 2002 Nov 22, towards the very end of this low state.  Although the
2-10~keV X-ray flux at the time is higher than observed earlier in the state, at 
$6\times10^{-12}$~erg~cm$^{-2}$~s$^{-1}$ it is comparable
to that observed by {\it Chandra} during the 2001 Feb low state
(U03).  {\it XMM-Newton} observed NGC~4051 for $\sim50$~ks with all
instruments, but due to background flaring only the central 33~ksec of
data was useful.  We therefore extracted EPIC-pn data (both single and
double pixel events) in this useful time range from
a 42~arcsec radius circle centred on the X-ray source, as well as a
source-free box outside the source region on the same chip for
background subtraction.  The EPIC instruments were operated using the
Medium filter and the EPIC-pn was operated in Large Window Mode
throughout the observation, and due to the faint nature of the source
pileup effects are negligible.  We do not use EPIC-MOS data in this analysis,
in order that we may directly compare the EPIC-pn data obtained during the low
state with that obtained during a normal state at an earlier
epoch before the recent change in EPIC-MOS response.  The EPIC-pn and
EPIC-MOS cross-calibration shows systematic discrepancies (up to 10\%)
between the instruments below 1~keV, so that it is unclear which gives the best representation of
the spectral shape at these low energies.  We note however that as a consistency check
we have also carried on the EPIC-MOS data the same spectral fits carried out on EPIC-pn
data in Section~\ref{bbmod}, and the results of
these fits do not change the basic conclusions we present
here (i.e. the same model is found to apply although specific model parameters
change slightly).  A detailed analysis of the RGS spectrum of NGC~4051, obtained during the {\it
XMM-Newton} observation of the low state is presented in a separate paper (Page et al., in prep.).

\section{Lightcurves and broadband spectra}
\label{lcs}
\begin{figure*}
\begin{center}
{\epsfxsize 0.7\hsize
 \leavevmode
 \epsffile{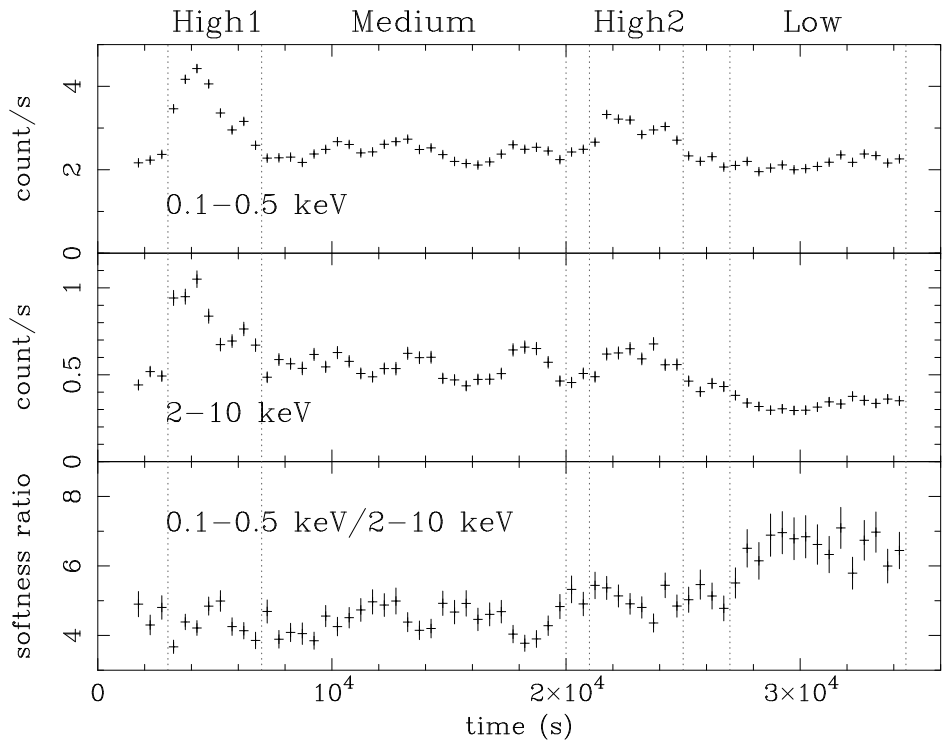}
}\caption{Soft and hard 500~s resolution light curves of NGC~4051, together with the
softness ratio between the two bands.  Dotted lines denote the low,
medium and both high flux epochs used to examine spectral variability in more
detail.} \label{shortlc}
\end{center}
\end{figure*}

\begin{figure*}
\begin{center}
{\epsfxsize 0.7\hsize
 \leavevmode
 \epsffile{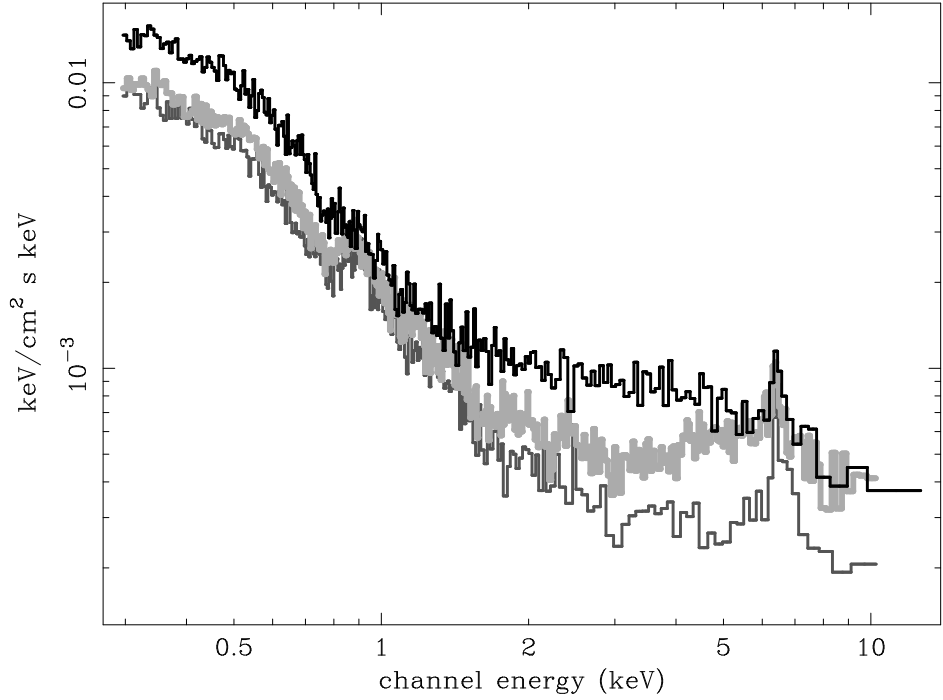}
}\caption{Unfolded broadband spectra (see Section~\ref{lcs} for
details) obtained during high1 (black line), medium (thick light grey line) and
low (dark grey line) flux epochs within the low state.  The spectra
are plotted in units of energy$\times$flux-density, so that a flat
slope corresponds to a photon index $\Gamma=1$.  For clarity
the spectra have been plotted as histograms and error bars omitted.
The spectra have been unfolded with respect to a simple power-law
absorbed by Galactic absorption (see text for details).} \label{spectra}
\end{center}
\end{figure*}

In Fig.~\ref{shortlc} we show the background-subtracted 
EPIC-pn light curves of NGC~4051 in two energy bands
(0.1-0.5~keV, which we shall call {\it soft} and 2-10~keV, which we
shall call {\it hard}).  We also show the varying softness ratio (soft/hard)
of the two bands.  The source shows clear flux variability and
interesting spectral variability during the low state.  In particular,
during the large `flare' early in the observation, the softness ratio
does not change substantially, and similar behavior can be seen during
a second weaker flare at $\sim22$~ks.  However, towards the end of the
observation (after $\sim26$~ks), 
when the source is fainter in 
both bands the source softens, as the hard flux decreases substantially
while the soft flux barely changes.  The source also hardens slightly
in reponse to small hard `flares' in the period between the two larger
flares.
In order to investigate the spectral variability in more detail, we
plot in Fig.~\ref{spectra} the unfolded (relative to a simple
power-law) EPIC-pn spectra corresponding to the
epochs delineated in Fig.~\ref{shortlc}, which correspond to three of the four
distinct periods discussed above, and which we refer to as high1,
medium, and low flux spectra.  For clarity, we do not plot the high2
spectrum, which is somewhat intermediate in shape between the high1 and
medium flux epoch spectra.  The spectra have been unfolded through the
instrument response with
respect to a simple power-law ($\Gamma=2$) absorbed by Galactic neutral
absorption ($N_{\rm H}=1.3\times10^{21}$~cm$^{-2}$,
\citealt{elv89}).  As noted in U03, although unfolded X-ray spectra are
always model-dependent, in the case of relatively high resolution CCD
instruments such as the EPIC-pn, only sharp features such as edges are
significantly misrepresented, while the general continuum shape is
fairly well represented by the unfolded spectra.
It can be seen immediately that the high flux epoch corresponds to an
increase in the soft flux, below $\sim0.7$~keV, with
little difference in the soft spectral shape between low, medium and high flux
epochs at those low energies.  Above $\sim2$~keV, the spectral shape
varies significantly.  Interestingly, in all epochs there is little change in the continuum
around $\sim1$~keV.  

Clearly the spectral variability cannot be
explained by any simple spectral model.  For example, the constancy in
flux around 1~keV might suggest some kind of pivot point around that
energy, however the soft spectral shape at lower energies appears to
be constant and is inconsistent with any low energy pivot.  Also, the
high and medium flux epoch spectra appear to converge above 5~keV,
despite both spectra showing quite different soft normalisations.  In
order to clarify the nature of the spectral variability, and place it
into context with the spectral variability seen at higher fluxes, we
now apply the flux-flux plot method to the data.

\section{Flux-flux plots}
\label{fluxflux}
\begin{figure*}
\begin{center}
{\epsfxsize 0.9\hsize
 \leavevmode
 \epsffile{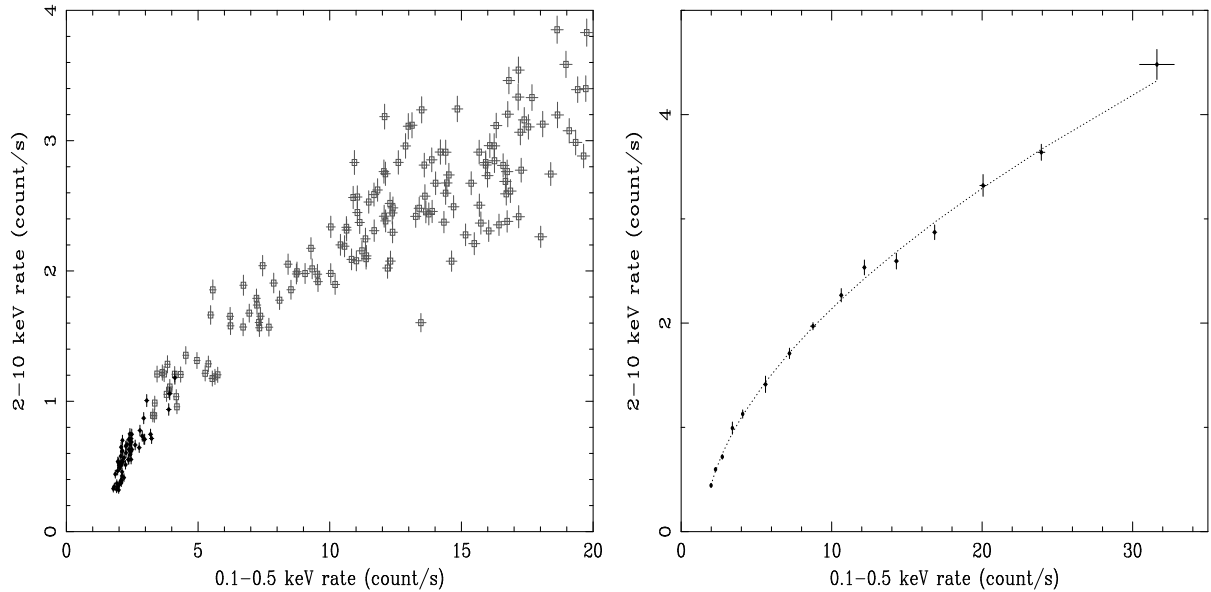}
}\caption{Left panel: Unbinned HS flux-flux relation.  Low state data
is plotted with filled markers, normal state data with lighter, open
markers.  For clarity only the 0-20 count/s soft flux range is
plotted, although all data (extending to 45 count/s) is included to
produced the binned relation (right panel). Right panel: Binned HS
flux-flux relation.  The dotted line shows the best-fitting power-law
plus constants model described in the text.} \label{hsfvfplots}
\end{center}
\end{figure*}
\citet{tay03} used a simple, model-independent technique, the
`flux-flux' plot, to examine the nature of the X-ray spectral variability of Seyfert
galaxies.  Specifically, by plotting the flux in a hard band versus
flux in a soft band, a linear relationship reveals that the data are
consistent with a simple two-component model, where one component
varies in flux but not in spectral shape, while the other component
remains constant in spectral shape and flux (see also \citealt{fab03},
\citealt{shi02}).  The spectral variability is then produced by a change in the relative
normalisations of the two components, with hardening at low fluxes
corresponding to a soft component varying in normalisation with respect to a constant
(or at most weakly-varying) hard component.  On the other hand, a
power-law flux-flux relationship implies spectral pivoting of the
varying component (so that the spectral variability is intrinsic to the
varying continuum).  

\begin{figure*}
\begin{center}
{\epsfxsize 0.9\hsize
 \leavevmode
 \epsffile{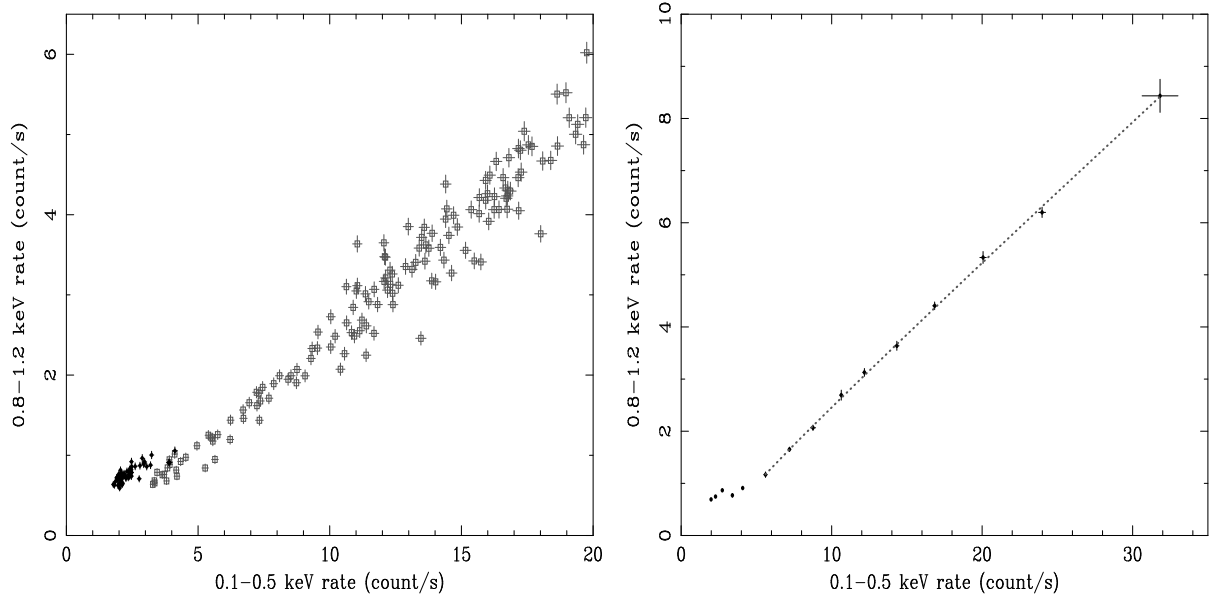}
}\caption{Left panel: Unbinned MS flux-flux relation.  Low state data
is plotted with filled markers, normal state data with lighter, open
markers.  For clarity only the 0-20 count/s soft flux range is
plotted, although all data (extending to 45 count/s) is included to
produced the binned relation (right panel). Right panel: Binned MS
flux-flux relation.  The dotted line shows the best-fitting power-law
plus constants model which fits the normal state data (see text for details).} \label{msfvfplots}
\end{center}
\end{figure*}

In order to examine the form of the continuum spectral variability of
NGC~4051, and test whether the low state spectrum is consistent with
that expected from the spectral variability seen at higher fluxes, we
can produce a combined flux-flux plot of EPIC-pn fluxes obtained
during the low state TOO observation and an earlier (2001 May)
$\sim100$~ksec {\it XMM-Newton} observation of NGC~4051,
obtained during its normal, higher flux state. 
The normal-state data is described in more detail
in \citet{mas02}, \citet{mch03} and see also \citet{sal03} and Salvi
et al. (in prep.) for an alternative spectral analysis of this data.
For both low and normal states, we
obtained 500~s resolution light curves in the 0.1-0.5~keV soft and
2-10~keV hard bands (see Section~\ref{lcs}), as well as an
additional 0.8-1.2~keV {\it medium} band.  Since the normal state EPIC-pn
observation was obtained in Small Window Mode, we renormalised the count
rates in both sets of light curves by livetime fraction, to account for
the different readout times used in the two observations.  We plot the
resulting hard {\it vs.} soft (HS) and medium {\it vs.} soft (MS) flux-flux
plots in Fig.~\ref{hsfvfplots} and Fig.~\ref{msfvfplots} respectively.  

\subsection{Hard {\it vs.} soft flux-flux relation}
We first examine the unbinned HS flux-flux relation (left panel in Fig.~\ref{hsfvfplots}).  We note
two main points.  First, the overall 
distribution of points clearly does not follow a linear relationship.
Second, the distribution of low state data seems to join smoothly on
to the distribution of normal state data: the normal state
distribution bends towards low fluxes and this bending continues in
the low state data.  Hence we conclude that the same process which causes
the spectral variability in the normal state continues to even lower fluxes
in the low state.  There is significant intrinsic scatter in the
relation (which was also observed in the flux-flux
relations plotted by \citealt{tay03}), which implies that there is
also spectral variability which is not correlated with flux
variations.  Note that this scatter increases towards higher fluxes
(see Appendix~\ref{app}). 
To remove the scatter and so find the functional form of the HS
flux-flux relation, we followed the method of \citet{tay03} and binned
the data into 15 flux bins (minimum of 10 data points
per bin), which in this case are logarithmically spaced in order to maximise the
usefulness of the low state data to constrain the flux-flux relation
(error bars are standard errors determined using the
spread of data points in each bin).  We use a general
power-law plus constants model (Taylor et al. 2003) to describe the
relationship of binned hard ($F_{h}$) and soft ($F_{s}$) fluxes:
 \begin{equation}
\label{eqn:fhfs1}
F_{h}=k(F_{s}-C_{s})^{\alpha}+C_{h}
\end{equation}
We find a good fit to the data ($\chi^{2}=10.8$ for 11 d.o.f.) for a power-law index
$\alpha=0.57\pm^{0.07}_{0.04}$ with constant offsets on the hard and
soft axes of $C_{h}=0.1\pm^{0.25}_{0.1}$ and $C_{s}=1.62\pm^{0.3}_{0.07}$
respectively\footnote{Unless otherwise noted, all errors
quoted in this paper represent 90\% confidence limits for one
interesting parameter, i.e. $\Delta \chi^{2}=2.71$.}. 
The binned HS flux-flux plot and best-fitting power-law
plus constants model are plotted in the right panel of
Figure~\ref{hsfvfplots}.  To demonstrate that this pivoting model is not simply a
result of the low-state and normal states following different, linear
relations, we also made and fitted a binned HS flux-flux relation for the
normal state data only (i.e. excluding the data points from the low
state TOO observation).  A linear plus constants model fit to the normal state HS flux-flux
relation is a very poor fit ($\chi^{2}=44$ for 11~d.o.f.), while a
much better fit is obtained with the power-law plus constants model 
($\chi^{2}=12$ for 9~d.o.f.), for power-law index $\alpha=0.63$,
consistent with our interpretation that the hard spectral shape in the
low state is a continuation of the spectral variability process seen
at higher fluxes.

Using {\sc xspec} simulations we find that the 90\%
confidence range of indices $\alpha$ corresponds to a pivoting model
with a pivot energy between 75-300~keV, consistent with the estimate
of 100~keV obtained from the {\it RXTE} data \citep{tay03}\footnote{We point out
here that the value of pivot energy of 300~keV quoted by us in Taylor
et al . (2003) is a typographical error.  A 100~keV pivot energy is in
fact a better match to the observed flux-flux relation slope, as can
be seen from Fig.~1 in Taylor et al. 2003}.  The positive offset on
the soft axis of the present flux-flux plot (which uses a lower soft
energy band than the 2-5~keV band used by Taylor et al.) implies a significant constant
soft component which counteracts the effect of the hard constant
component detected at higher energies \citep{tay03} to produce an offset on the soft axis.

\subsection{Medium {\it vs.} soft flux-flux relation}
We now consider the unbinned MS flux-flux relation (left panel in
Fig.~\ref{msfvfplots}, and see right panel for the binned up
relation).  The relation appears close
to linear during the normal state but seems to flatten off in the low
state (corresponding to the weak variability in the medium band seen in
Fig.~\ref{spectra}).  Binning up and fitting only the normal state data (i.e. the
nearly linear portion of the plot) with Equation~\ref{eqn:fhfs1} used above, we
obtain a good fit ($\chi^{2}=5.9$ for 9 d.o.f.) for an
index $\alpha=0.97\pm0.06$.  Simulations show that if 
we assume the 75-300~keV range of acceptable pivot energies implied by
the fit to the HS flux-flux relation, the slope of the MS relation should be
0.78-0.82.
 Therefore it is likely that some other varying component (in addition to
the pivoting power-law), is present in both the medium and soft bands, in
order to wash out the effect of pivoting between the two bands.  The
simplest possibility, implied by the apparent lack of
spectral-shape variability below 0.7~keV (Fig.~\ref{spectra}) is that
this component has a constant spectral shape.  The shape of the soft
excess flattens towards low energies, which suggests that the additional
soft spectral component in NGC~4051 is thermal, rather than a power-law
(\citealt{col01}, U03), so one possible configuration is that the pivoting power-law
is attached to a variable soft thermal component of constant shape
(although note that \citealt{ogl03}, and Salvi et al., in prep., explain this flattening in the soft
excess by invoking an extreme relativistic O{\sc viii} diskline in addition
to a steep power-law continuum).  One might
then ask why the MS flux-flux relation flattens at low fluxes.  As
discussed by \citet{tay03}, the gradient of a flux-flux plot for a
variable spectral
component with a constant spectral shape is equivalent to the hardness ratio of that
component.  Therefore the flattening of the flux-flux plot at low fluxes
can be caused by a strong softening of the varying spectral component
in the energy range covered by the
plot.  Such an effect could be caused if the variable component of emission is
largely removed by absorption in the 0.8-1.2~keV band, but not in
0.1-0.5~keV (e.g. if the absorber is ionised). 
This possibility is consistent with
the spectra in Fig.~\ref{spectra}: the high flux spectrum appears to
converge with the lower flux spectra above $\sim0.7$~keV, an effect which could
be caused by absorption, if the expected edges (e.g. due to O{\sc
vii}) are partly filled in by a constant, unabsorbed soft continuum component.  The fact
that the spectral shape below 0.5~keV appears not to vary with flux
(see Fig.~\ref{spectra}) suggests that the constant and variable soft
components have a similar spectral shape, at least at low energies.

\section{Spectral modelling}
\label{bbmod}
We now apply the physical interpretation of the spectral variability
of NGC~4051 inferred from the flux-flux plots to construct a broadband model which
can fit the flux-dependent spectra shown in Fig.~\ref{spectra}.
First however, to complete our picture of the appropriate spectral components for our broadband
spectral fit, we examine the spectrum (obtained over the entire useful
exposure) in the 3-12~keV band.   
\begin{figure}
\begin{center}
{\epsfxsize 0.9\hsize
 \leavevmode
 \epsffile{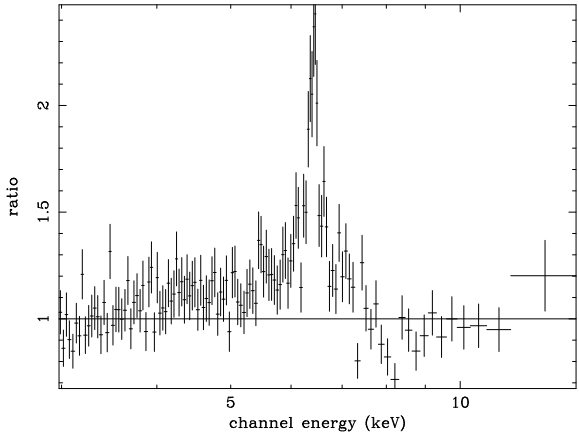}
}\caption{Ratio of 3-12~keV EPIC-pn data to a simple power-law fitted
over that range but excluding data over 4-9~keV.} \label{hardspec}
\end{center}
\end{figure}
\subsection{Hard spectrum}
\label{hardfit}
In Fig.~\ref{hardspec} we plot the
ratio of the 3-12~keV spectrum to a simple power-law model, which is fitted
only over the 3-4~keV and 9-12~keV ranges (i.e. excluding the region
where significant emission features may be expected).  Clear broad
residuals can be seen, in addition to a narrow iron emission line observed
at $\sim6.4$~keV.  Consequently,
a power-law plus narrow, unresolved Gaussian fitted to the data is a very poor fit
(reduced chi-squared, $\chi^{2}_{\nu}=1.94$ for 120 d.o.f.).  
We find that adding a {\sc pexrav}
reflection component\footnote{Similar to the procedure in U03, 
apart from reflection scaling factor $R$, which
is free, reflection model parameters are fixed
to those corresponding to the values from the best-fitting pure
reflection model used in \citet{utt99}, i.e. cut-off energy 100~keV,
inclination angle 30$^{\circ}$, fixed illuminating
power-law photon index  $\Gamma=2.3$ and normalisation at 1~keV
$A=0.01$~photon~cm$^{-2}$~s$^{-1}$~keV$^{-1}$.  
Leaving these parameters free does not improve
the fit significantly.} to the model improves the fit substantially
($R=2.4\pm^{0.5}_{0.7}$), 
and an edge at $7.9\pm{0.15}$~keV (optical depth $\tau=0.5\pm^{0.25}_{0.15}$) 
is also formally required (resulting $\chi^{2}_{\nu}=0.91$ for 117
d.o.f).  The observed edge energy corresponds to the K-shell photoionisation
edge expected from Fe{\sc xvii}.
Removing the edge and including a Laor diskline (as suggested by the
spectral modelling of U03), worsens the fit ($\Delta \chi^{2}=+13$ for
one less degree of freedom).  Adding the 7.9~keV edge in
addition to the diskline does not improve on the reflection+edge fit
significantly ($\Delta \chi^{2}=-3$ for three fewer degrees of
freedom).  Worse fits are obtained in each case by including a
diskline from a non-rotating black hole.
Therefore we conclude that a diskline is not formally required to fit
the low state spectrum, and so for the purposes of our broadband spectral fit we will use the
simple reflection plus edge model to account for
the broad residuals in the hard spectrum.  Note however that we regard this
model for the hard spectral shape as a phenomenological and not
necessarily a physically motivated representation of the hard spectrum (see the Discussion
for further details of the model interpretation).  The narrow iron line
($E_{\rm Fe}=6.42\pm0.01$~keV) contributes a flux of
$(1.3\pm0.2)\times10^{-5}$~photon~cm$^{-2}$~s$^{-1}$, consistent with
the flux observed by {\it Chandra} in 2000 April \citep{col01} and
2001 February (U03).  It is likely that this line is constant and originates some distance from
the continuum source, as implied by the narrow velocity width observed
by the {\it Chandra} HETG \citep{col01}.  

\subsection{The model}
We now apply the information gained from the flux-flux plots, together
with the model for hard spectral features, to construct our broadband spectral
model.  Our {\sc xspec} model is: {\sc phabs} $\times$ ({\sc
zedge} $\times$ {\sc zedge} $\times$ {\sc zedge} $\times$ ({\sc pegpwrlw} +
{\sc pexrav} + {\sc
compbb}) + {\sc compbb} + {\sc zgauss} + {\sc zgauss} + {\sc zgauss} +
{\sc zgauss} + {\sc zgauss} + {\sc zgauss}).  Note that although the
model appears to be complicated, many of the model parameters are fixed (for
example, all the Gaussian line parameters).  Furthermore, the same
model is used to simultaneously fit the spectra from all four flux epochs within the
low state, with only a small subset of parameters left free to vary
between the spectra, and most parameters forced to find the same best
fitting value for all four spectra (see descriptions below).   We now
describe the model components, and the rationale for including them:
\begin{enumerate}
\item Galactic neutral absorption ($N_{\rm H}$ fixed at
$1.31\times10^{20}$~cm$^{-2}$, \citealt{elv89}) and edges fixed (in the source rest
frame, $z=0.0023$) at 0.74~keV 
(O{\sc vii}), 0.87~keV (O{\sc viii}) and 7.9~keV (probable Fe{\sc
xvii}).  The 7.9 keV edge is required by the fit to the hard spectrum
and we include the two oxygen edges (which are also suggested by
\citet{col01}, to fit a {\it Chandra} grating
observation of NGC~4051 at higher fluxes), to
represent (in simple terms) the ionised absorber which may be required
to explain the shape of the MS flux-flux relation at low fluxes.  The optical depths
of the edges are free, but forced to be identical between spectra.
Note that other absorption features, such as the unresolved transition array
(UTA) of Fe (Page et al. in prep.) may also contribute significantly to absorption above
$\sim0.7$~keV, but for simplicity we only model absorption in terms of
the O edges.
\item A pivoting power-law ({\sc xspec} model {\sc pegpwrlw}),
with pivot energy fixed to 100~keV and normalisation at the pivot
energy fixed to 0.078 $\mu$Jy (determined from the {\it RXTE}
flux-flux plot, Taylor et al. 2003).  This component is required to
simply explain the shape of the HS flux-flux plot, and the flux-flux
relation at higher energies reported by \citet{tay03}. 
\item A reflection component ({\sc pexrav}) with parameters fixed to those used in
Section~\ref{hardfit} and normalisation fixed to be the same in all
fitted spectra (i.e. to approximate the constant hard component
suggested by Taylor et al. 2003).
\item A {\it constant} Comptonised blackbody which is {\it unabsorbed} by
the oxygen edges.  This
component is used to represent the soft constant component which is
required by the flux-flux relations.  The normalisation, blackbody
temperature, electron temperature and optical depth are free but
forced to be identical between spectra.  Note that a Comptonised blackbody is
the best fitting model we have found to describe the soft component.  See Section~\ref{fitres} for
details of fits using other models to describe the soft emission.
\item A {\it variable} Comptonised blackbody which is {\it absorbed} by the oxygen
edges.  This component represents the variable soft component of
constant shape which can explain the form of the MS flux-flux
relation, together with the apparent lack of spectral variability
below 0.7~keV.  The lack of spectral variability at low energies
also suggests that the variable soft component has a similar shape to
the constant soft component, so we allow the normalisation of the
variable Comptonised blackbody component to be free between spectra
while forcing the blackbody temperature, electron temperature and
optical depth to be the same as for the constant Comptonised blackbody
(component iv).
\item A narrow 6.4~keV iron K$\alpha$ line (unresolved Gaussian,
fixed flux $1.3\times10^{-5}$~photon~cm$^{-2}$~s$^{-1}$), to represent
the line observed in the hard spectrum.  Additional
prominent soft emission lines (e.g. see \citealt{col01})
(as unresolved Gaussians), with fixed fluxes are required to represent the
most prominent features observed in the RGS spectrum
(fluxes fixed at those observed in the RGS spectrum, Page et al. in prep.): 0.739~keV O{\sc vii} RRC
($1.9\times10^{-5}$~photon~cm$^{-2}$~s$^{-1}$), 0.725~keV Fe{\sc xvii}
($4.1\times10^{-5}$~photon~cm$^{-2}$~s$^{-1}$), 0.654~keV O{\sc viii}
Ly~$\alpha$ ($7.2\times10^{-5}$~photon~cm$^{-2}$~s$^{-1}$), 0.56~keV
blend of O{\sc vii} x+y, z
($7.2\times10^{-5}$~photon~cm$^{-2}$~s$^{-1}$).  A number of lines are
also seen in the RGS spectrum around 0.9~keV, including Ne{\sc ix} and
the O{\sc viii} RRC, however these lines alone do not replicate a
broad bump in the spectrum around 0.9~keV, which may possibly be a
blend of iron emission or an unusual continuum shape
(see Page et al., in prep.).  Therefore we
include a broad Gaussian ($\sigma=0.05$~keV, flux
$1.52\times10^{-4}$~photon~cm$^{-2}$~s$^{-1}$) at 0.91~keV to crudely represent
these combined features.
\end{enumerate}
The model is fitted simultaneously to all the spectra from low, medium and
both high flux epochs, so that in total there are 16 free parameters.
To reiterate, these free parameters are the
optical depths of the three edges (fixed to be identical between
spectra), the temperature, electron temperature, and optical depth of
the Comptonised blackbodies and the normalisation of the constant
Comptonised blackbody (also fixed to be identical between
spectra), the normalisation (here we use the scaling factor $R$) of
the constant reflection component, four different normalisations (between spectra) of the
variable Comptonised blackbody and four different indices
(between spectra) of the pivoting power-law.
\begin{figure*}
\begin{center}
{\epsfxsize 0.7\hsize
 \leavevmode
 \epsffile{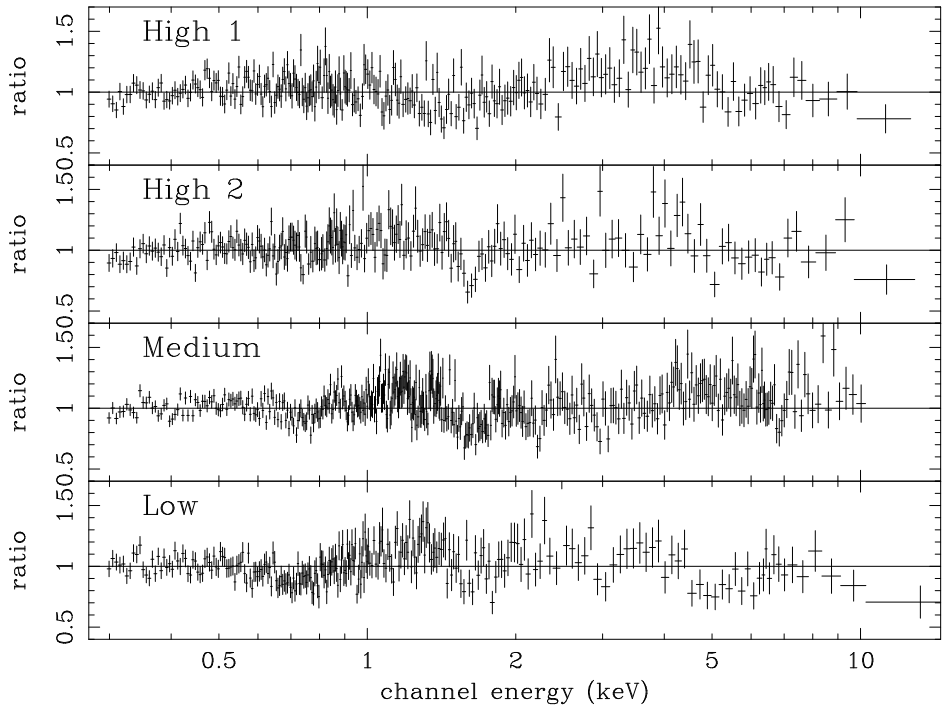}
}\caption{Ratio plots of the data to the best-fitting model (described in the text) to
for low, medium and both high flux epochs in the low state.} \label{modratios}
\end{center}
\end{figure*}
\begin{figure*}
 \par\centerline{\psfig{figure=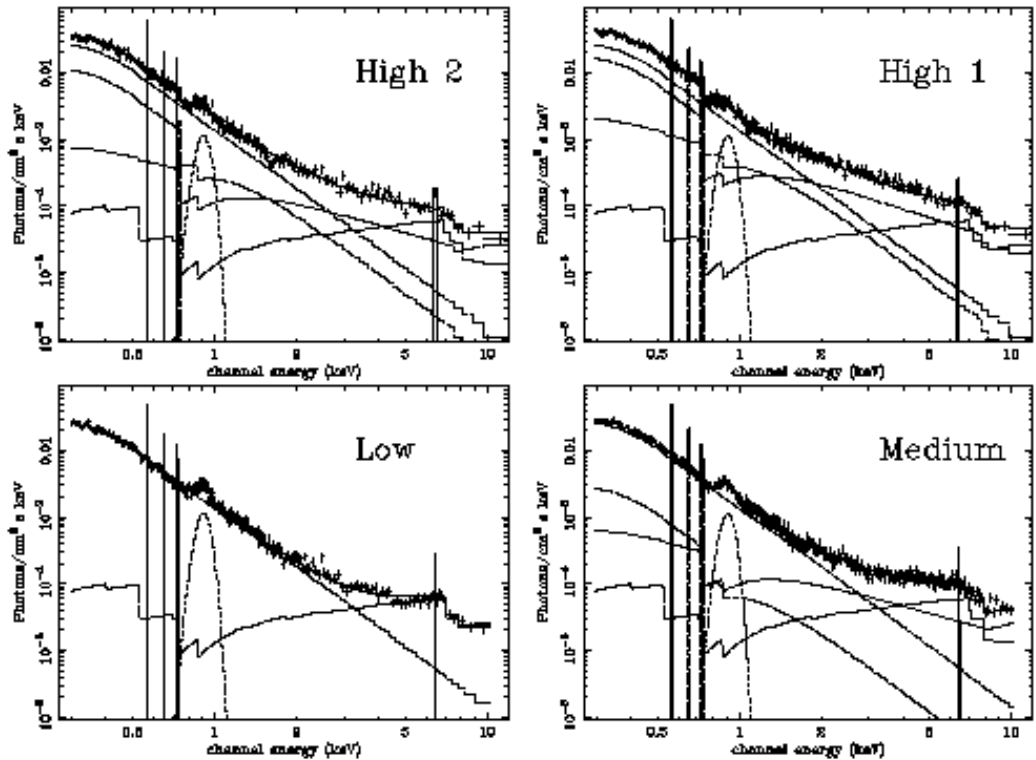,width=14cm}}
 \caption{\label{ufmodspec} 
Unfolded EPIC-pn spectra for low, medium and both high-flux
epochs during the low state.  The various model components are plotted
with lines.  Note that power-law and absorbed Comptonised
blackbody components do not contribute significantly to the low flux
epoch spectrum.
   }
\end{figure*}

\subsection{Fit results}
\label{fitres}
The best fitting model yields $\chi^{2}_{\nu}=1.37$ for 1159 d.o.f.,
for model parameters which are shown in Table~\ref{fitpars}.  Data/model
ratios are plotted in Fig.~\ref{modratios}.  Significant residuals
still exist (see the Discussion for consideration of the causes
of these residuals), although the model can reproduce the general
spectral shape and variability reasonably well.  Note that the
apparent `absorption trough' observed at $\sim1.6$~keV 
in the second high flux epoch spectrum does not correspond to
the probable weak systematic feature in the detector response 
observed at $\sim1.8$~keV in the EPIC-pn spectrum of
MCG--6-30-15 \citep{fab03}.  However, since this 1.6~keV feature in
the EPIC-pn spectrum of NGC~4051 is not replicated in the EPIC-MOS
spectra from this epoch it is likely to be a statistical
artefact.  Although the
fit is not formally acceptable, the model appears to give a reasonable
description of the overall spectral variability. 
\begin{figure*}
\begin{center}
{\epsfxsize 0.6\hsize
 \leavevmode
 \epsffile{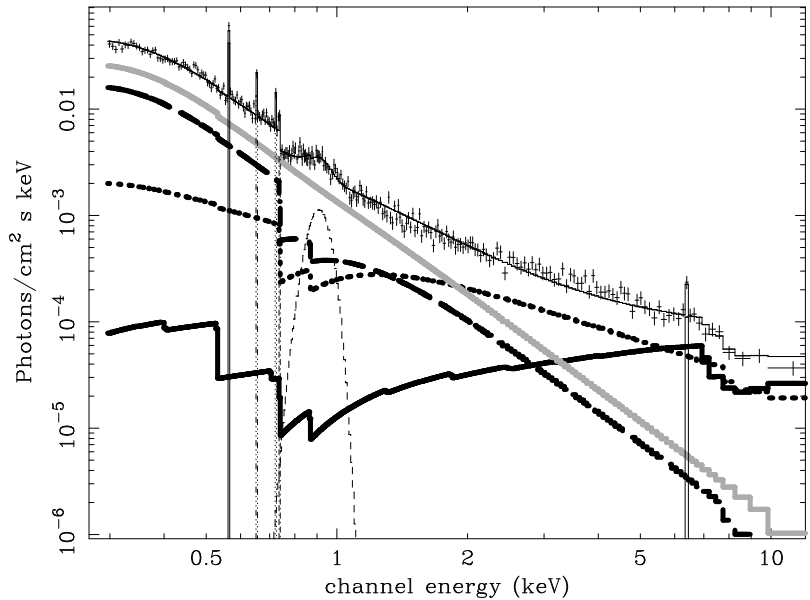}
}\caption{Unfolded model fit to the high1 spectrum.  To make the
contributions of the different continuum components clearer, they are
plotted using thick lines.  The absorbed and unabsorbed Comptonised
blackbody components are plotted with a black dashed line and a solid
grey line respectively.  The reflection component and the absorbed
pivoting power-law component are plotted with a solid black line and a
black dotted line respectively.} \label{ufhigh1}
\end{center}
\end{figure*}
To demonstrate the effects of the different model components on the
spectral variability, we plot the unfolded models and spectra for the
low, medium and both high flux epochs in Fig.~\ref{ufmodspec}.  A more
detailed plot of the unfolded model fit to the high1 spectrum is also
shown in Fig.~\ref{ufhigh1}, with the various continuum components
made more distinct to allow easier interpretation.
We note that the model
can explain the complicated spectral variability reported in
Section~\ref{lcs} as being largely due to the variation of the hard and
soft variable components relative to the constant components in the
respective bands.  In other words, the effect of power-law pivoting dominates
the overall spectral variability of the source (i.e. across the entire
observed flux range, as revealed in Fig.~\ref{hsfvfplots}), but within the
low state where the varying continuum flux is small,
spectral variability due to the presence of constant
components is much more important. 
For example, the hardening in response to small
flux increases occurs when the soft variable thermal component is very
weak compared to the constant soft component,
so that the flux variation is primarily in the hard band,
associated with the pivoting power-law (which though weak is still significant
compared with the constant reflection component in that band).  
On the other hand, a large flux increase (e.g. the flare early in the observation)
corresponds to the flux of the variable thermal component also increasing
significantly, in addition to the hard power-law flux, so that the
soft/hard flux ratio does not change. 
Note also that the divergence of the medium and high flux
spectra below $\sim5$~keV is explained by the model as being due to
the larger Comptonised thermal component in the high flux spectrum.
At the lowest fluxes the power-law contribution to the EPIC-pn
spectrum is negligible, i.e. the spectrum is probably
dominated by the constant thermal and reflection components.

\begin{table*}
 \caption{Best-fitting broadband model parameters.}
 \label{fitpars}
 \begin{tabular}{@{}lcccccccccc}
     & $\Gamma$ & $A_{\rm cbb}$ & $A_{\rm const}^{f}$ & $kT_{\rm cbb}^{f}$ &
$kT_{\rm e}^{f}$ & $\tau_{\rm cbb}^{f}$ & $\tau_{\rm O{\sc vii}}^{f}$ &
$\tau_{\rm O{\sc viii}}^{f}$ & $\tau_{\rm Fe}^{f}$ & $R^{f}$ \\
 Low & -2.5 & 0.0 & 4.02 & 0.08 & 18.0 & 1.08 & 1.3 & 0.5 & 0.4  & 2.4 \\
\\
 Medium & 1.15 & 0.40 & 4.02 & 0.08 & 18.0 & 1.08 & 1.3 & 0.5 & 0.4 & 2.4 \\
\\
 High 2 & 1.17 & 1.68 & 4.02 & 0.08 & 18.0 & 1.08 & 1.3 & 0.5 & 0.4 & 2.4 \\
\\
 High 1 & 1.35  & 2.52 & 4.02 & 0.08 & 18.0 & 1.08 & 1.3 & 0.5 & 0.4  &
2.4 \\
 \end{tabular}

\medskip
$\Gamma$ is the photon index of the pivoting power-law; $A_{\rm cbb}$
and $A_{\rm const}$ are the normalisations of the variable and
constant Comptonised blackbodies respectively (units of $10^{4}$); $kT_{\rm cbb}$,
$kT_{\rm e}$ and $\tau_{\rm cbb}$ are the blackbody temperature (in keV),
and Comptonising electron temperature (in keV) and optical depth for the
Comptonised blackbodies; $\tau_{\rm O{\sc vii}}$, $\tau_{\rm O{\sc
viii}}$ and $\tau_{\rm Fe}$ are the edge optical depths and $R$ is
the reflection scaling factor of the constant reflection component
(see Section~\ref{hardspec}).  Note that since the fit is not formally
acceptable, meaningful errors cannot be quoted.\\
$^{f}$ Values forced to be the same for all four spectra.
\end{table*}

For completeness, we also substituted different soft spectral models
for the constant and variable soft components used in the fit.
Subsituting the Comptonised blackbody spectra with
a {\sc diskpn} model \citep{gie99} for a pure disk blackbody spectrum
modified by a pseudo-Newtonian potential (to take account of
relativistic effects), we obtain a much worse fit ($\chi^{2}_{\nu}=3.1$
for 1160 d.o.f.).  If we instead substitute simple steep power-laws,
the fit is poorer ($\chi^{2}_{\nu}=1.52$ for 1161 d.o.f., for photon
index $\Gamma\simeq3$) than obtained using Comptonised blackbodies,
with the main residuals corresponding to a flattening of the observed
spectrum below 0.5~keV.  We conclude that, of the simple models tested
here the Comptonised blackbody provides the best description of the
soft component spectra (although see \citealt{ogl03}, \citealt{sal03} and Salvi et al.,
in prep., for an alternative description invoking a relativistic
O{\sc viii} diskline and a steep power-law).

\section{Discussion} 
We have demonstrated that the information gained from flux-flux plots
can be used to infer the form of a spectral model which is relatively
successful at replicating the seemingly complex X-ray spectral behaviour of NGC~4051
observed in the low state.  Of course, significant residuals still
remain in our model fits, and so our model should be considered only
as an approximation to the true spectral behaviour.  In this section,
we will discuss some of the implications of our results and describe
some of the practical and theoretical aspects of improving our model in relation
to these implications.

\subsection{The variable continuum components}
The flux-binned 0.1-0.5~keV {\it vs.} 2-10~keV flux relation is well
fitted by a power-law plus constant model, implying that the relation
between these two bands, and the bulk of the spectral variability
observed in NGC~4051 (over its entire flux range), is
caused by the spectral pivoting of the power-law, also observed in the 2-15~keV band
\citep{tay03}.  The presence of a constant-shape thermal
component (which we model here with a Comptonised blackbody), which
dominates the continuum at low energies, suggests that the pivoting
power-law must be `attached' to the thermal component, thus producing
the observed flux-flux relation and the good correlation observed
between extreme UV and medium energy X-ray fluxes  \citep{utt00}.  An
obvious physical realisation of this model is that the thermal
component represents a {\it directly observed} variable source of
seed-photons, which are upscattered in the corona to produce the
variable power-law emission.  In fact the observed blackbody
temperature is close to the $\sim0.1$~keV temperature expected from the small black
hole mass of $\sim3\times10^{5}$~M$_{\odot}$, inferred for NGC~4051
from both reverberation mapping \citep{she03} and the X-ray
variability power spectrum \citep{mch03}.  The spectrum of NGC~4051 may then be
unusual in this regard, since most AGN have larger black hole masses
and hence their disk blackbody emission will not be directly observable in
the soft X-ray band.  

The fact that NGC~4051 appears to show a very strongly
varying thermal component indicates a highly unstable accretion
disk.  Recent interpretations of AGN X-ray variability in terms of
fluctuations in the accretion flow (implied by the observed `rms-flux'
relation and the similarity with X-ray binary systems,
\citealt{utt01,utt03b}) suggest that the accretion flow is geometrically
thick, so that perturbations can propagate from a range of radii into the
inner X-ray emitting regions.  For AGN variability dominated by intrinsic variability of the
power-law emission and not varying thermal emission, a geometrically
thick but optically thin coronal
accretion flow or advection dominated flow may suffice to explain the variability.  However
in the case of NGC~4051, where the thermal emission appears to
drive the pivoting of the power-law (see below), the optically thick
accretion disk must itself be geometrically thick in order for the
model of propagating perturbations in the accretion flow to work
(e.g. see Appendix to \citealt{chu01}).
This situation can be realised at high accretion rates (e.g. \citealt{fra92}).  The long-term
average 2-10~keV X-ray luminosity of NGC~4051 from {\it RXTE}
monitoring (e.g. \citealt{mch03}) of $\sim3\times10^{41}$~erg~s$^{-1}$,
corresponds to about 1\% of the
Eddington luminosity for the black hole mass given above.  Depending on the
assumed bolometric correction, the total luminosity is likely to be of
the order of a few tens of \% of Eddington, possibly sometimes
reaching close to the Eddington rate, so a thick disk seems a likely
configuration for the accretion flow in NGC~4051.

The physical connection between the thermal and power-law components
will determine the shape of the spectrum in the intermediate energy
range (perhaps 0.5-3~keV) where both components contribute
significantly.  For example, we might expect a rather smooth join
between the components, with a shape determined by the underlying thermal spectrum and the
temperature distribution of the Comptonising electrons, as well as
other factors (e.g. geometry of the emitting regions).  Therefore it
is not surprising that our rather crude best-fitting model, which treats both components
separately, is still not a very good fit to the data, especially in
that intermediate energy range, where notable residuals can be seen.  
Additional complexities might exist,
for example are there two populations of Comptonising electrons (as
suggested by our model which includes both a hard pivoting power-law and a
steeper Compton tail to the thermal component), or do the electron
populations merge (e.g. as in a hybrid thermal-non-thermal plasma,
\citealt{gie99})?  Spectral modelling with a more detailed
theoretical underpinning may help answer these questions by providing
much better fits to the data.

The pivoting power-law can be interpreted simply in terms of
models where the luminosity of the X-ray emitting corona is constant
and seed photons are varying (see \citealt{zdz02} and discussion in
\citealt{tay03}).  However, we note that there is intrinsic scatter
in the unbinned flux-flux plots, implying an additional source of
spectral variability which is independent of flux.  
We note that various spectral timing properties of NGC~4051 imply an
additional variability process in the hard band which applies on short
time-scales only (e.g. the
flattening of the power spectrum at higher energies and
the decrease in coherence between bands on short time-scales;
\citealt{mch03}), which is likely to be the origin of much of the
scatter in the HS flux-flux plot. 
\citet{tay03} suggested that the intrinsic scatter in the flux-flux plots
may be due to weak variability of the hard component (possibly
reflection) which is uncorrelated with the strongly variable power-law
component.  However, the fact that the scatter increases towards
higher fluxes, as is shown in Appendix~\ref{app}, implies that the
additional variable process is `aware' of the large amplitude flux variability and hence is
likely to be intrinsic to the variable power-law (note that this effect is
related to the rms-flux relation observed in NGC~4051, \citealt{mch03}).  Such additional 
variability in the power-law could arise from events which heat the corona on
short-time-scales (e.g. magnetic flares) independent of the
variations in seed photon flux which may be due to accretion instabilities.

\subsection{Reflection, absorption and the nature of the
constant continuum components}
A strong reflection component (which we assume is constant)
seems to be required to explain the data.  The normalisation of the
reflection component corresponds to a typical illuminating continuum
2-10~keV flux of $\sim3\times10^{-11}$~erg~cm$^{-2}$~s$^{-1}$ for a
reflector covering $2\pi$ steradian solid angle, as seen from the illuminating source
(compare with the {\it RXTE}-observed average 2-10~keV flux since 1996
of $2.2\times10^{-11}$~erg~cm$^{-2}$~s$^{-1}$).  One might assume
therefore that the constant reflection could originate from a distant
reflector.  However, the reflection required to fit the 2002 low state spectrum
is a factor$>2$ {\it larger}
than than that used to explain the May 1998 spectrum observed by {\it
BeppoSAX} and {\it RXTE} \citep{utt99}, 
while the narrow 6.4~keV line flux is {\it not larger} than observed
in May~1998, suggesting that a significant component of the
reflection is not directly coupled to the narrow line flux and may not arise in a
distant reprocessor.  Furthermore,
we point out that according to Fig.~\ref{longlc}, at earlier epochs in the 2002
low state the 2-10~keV source flux dropped to significantly lower levels than
we observed with {\it XMM-Newton}.  Since our spectral fits suggest
the source to be reflection dominated at the lowest fluxes observed by
{\it XMM-Newton}, this suggests that the some significant fraction of
the reflection component may
itself vary on time-scales of weeks or less.  

Weak variability of a reflection
component, which is independent of the illuminating continuum flux is
also implied by fitting a two-component model to data from
MCG--6-30-15 \citep{fab03}.  Puzzlingly however, a
diskline appears not to be required in NGC~4051, perhaps suggesting
that more realistic models than we use here are required to explain
the reflection,
i.e. taking full account of relativistic smearing, disk ionisation or more complex reflection
geometries (e.g. \citealt{fab02}).  We note here that 
\citet{min03b} demonstrate in the specific case of
NGC~4051 how strong, almost constant reflection can be produced by the
effects of strong gravitational light bending on X-rays illuminating
the disk from a source of variable height close to the black hole.
\citet{min03b} point out that in low flux states (when the source is
closest to the black hole, and hence light bending and relative
reflection is most extreme), the inner disk is likely to be highly ionised,
with attendant effects on the line strength and profile.
Indeed the extreme variability of the
thermal component in the spectrum of NGC~4051 (which may drive the
spectral pivoting) further suggests that the inner accretion disk in NGC~4051 is very
unstable, which may carry implications for the formation of prominent
iron disklines. 

We also detect a strong ($\tau\sim0.4$) absorption edge at $\sim7.9$~keV, possibly due
to Fe{\sc xvii}.  The presence of such a strong edge is puzzling,
because it corresponds to an absorbing column of Fe
$>10^{19}$~cm$^{-2}$, i.e. a hydrogen column $>10^{23}$~cm$^{-2}$
assuming Solar abundances.  In the absence of a large overabundance of
iron, it is difficult to imagine how such a large absorbing column
could remain hidden at lower energies.  This is because even if it is highly
ionised (e.g. ionisation parameter $\xi
\sim 10^{3}$--$10^{4}$), such a large column of gas would produce
strong edges in lower-$Z$ elements
which are not observed in the {\it XMM-Newton}
grating spectrum of NGC~4051 in the normal state
\citep{ogl03}.  One cannot also simply argue that the column
of gas in the line of sight is much larger in the low state, because
the resulting discontinuity in the effects of absorption
would not produce the gentle bending of the HS flux-flux relation
between the normal and low state seen in Fig.~\ref{hsfvfplots}.  Note
that these arguments appear to rule out the recent partial ionised absorber
explanation for the spectral variability of NGC~4051 suggested by
\citet{pou03}, but these arguments also impose strong constraints on our
interpretation of the edge at 7.9~keV detected in the low state.  
An alternative interpretation of the 7.9~keV edge is that it originates in
reflection, rather than absorption, i.e. it may originate from the
possibly constant reflection component.  This interpretation would
help to explain why this edge is not readily apparent in the normal
state spectra (e.g. it is not reported by \citealt{lam03a}), since it would be
diluted by the higher continuum level in that state.  Unfortunately,
the spectra from different flux epochs within the low state do not have sufficient high
energy signal-to-noise to test the flux dependence of the depth of the
edge, but our hypothesis at least appears to be
consistent with the lack of evidence for a very large column of
absorbing gas at lower energies.  

The presence of an absorption edge
of Fe{\sc xvii} in reflection is consistent with a mildly ionised reflector with
$\xi\sim100$.  Such a reflector would also produce edges of O{\sc vii}
and O{\sc viii} and it is tempting to equate these features with the
strong absorption edges ($\tau\sim1$) we use to fit the low state
data, which are required by the model in conjunction with a constant
soft continuum component,
in order to explain the lack of variability above $\sim0.7$~keV and
the resulting flattening of the MS flux-flux relation at low fluxes.
In fact, one might explain both the soft absorption features and the
constant soft component in terms of reflection of the primary
continuum (in this case a Comptonised
blackbody plus power-law), from different
locations of an ionised disk.  For example, if at low fluxes (where
the constant, unabsorbed soft component dominates) only the
innermost, heavily ionised regions of a disk are illuminated,
then the reflected spectrum will be almost featureless and match
closely the illuminating soft spectral shape.  However, if the
geometry of the source changes, so that at slightly
higher fluxes the source height increases or the source becomes more
vertically extended, the outer, less heavily ionised regions of the disk will
also be illuminated and edges will be imprinted on the reflected
spectrum.  Or alternatively, at higher illuminating fluxes the outer disk
may become sufficiently ionised to allow
reflection in soft X-rays together with the accompanying edges,
without requiring a change in source geometry.
At even higher fluxes, those edges may be reduced by further
ionisation (and hence do not appear in the normal state) or they may be
diluted if the soft component of reflection somehow remains constant at higher
fluxes, which is
possible if the soft reflection has the same origin as the constant
reflection component observed at harder energies.  
The basic physical picture we propose above
to explain the constant soft component and possible edges may be
consistent with the model of \citet{min03b}, if it is extended to treat
reflection from an ionised disk.  However, it is unclear whether our
interpretation that the variable thermal emission drives the power-law
pivoting could apply in the model of \citet{min03b}, since in that
model the continuum variability is explained as being predominantly
due to gravitational effects coupled with variations in height of a constant source.

An alternative explanation of the O edges (and/or possible Fe UTA)
required by our model is that they represent simple absorption of the
varying component of emission.  In this case, the constant soft
component is unabsorbed.  This picture represents something like the
partial covering scenario envisaged by \citet{pou03}, albeit without
the requirement of such a large column denisty of material, which seems
problematic (as noted above).  The optical depth of O edges required
by our model correspond to ionic column densities of $\sim10^{19}$~cm$^{-2}$
which are not easy to reconcile with the column densities observed in
grating data of the normal state ($\sim2\times10^{17}$~cm$^{-2}$, \citealt{ogl03}), even
given a large drop in ionisation parameter.  However the
predicted total column in this case is $\sim10^{22}$~cm$^{-2}$, lower
than the value estimated by \citet{pou03}, and
may be consistent with the lack of strong edges observed at higher
fluxes in the normal state.  The requirement of only partial covering in this scenario remains
difficult to explain however.

\section{Conclusions}
We have presented an analysis of EPIC-pn data from an
{\it XMM-Newton} observation of NGC~4051 in the low state in 2002
November.  We reach the following conclusions:
\begin{enumerate}
\item The spectral variability observed in
NGC~4051 in the low state is complex (see Section~\ref{lcs}).  However, flux-flux plots
show that the overall spectral shape in the low state is consistent
with an extrapolation of the spectral pivoting of a
power-law continuum observed at higher
fluxes (Section~\ref{fluxflux}), confirming the interpretation of
earlier {\it Chandra} data by Uttley et al. (2003).
\item At softer energies, the variable spectrum is reasonably well described by
a thermal component with a constant, Comptonised blackbody shape,
which in order to explain the spectral variability, may be
absorbed in the $\sim0.7$--1~keV range, perhaps by
photoelectric edges of O{\sc vii}/O{\sc viii} and/or unresolved
transition arrays of Fe, which may possibly be associated with
reflection features also seen at higher energies.  The blackbody
temperature ($kT\sim0.1$~keV) of this Comptonised thermal component is
consistent with the low black hole mass of NGC~4051
($3\times10^{5}$~M$_{\odot}$, \citealt{she03}), and the strong
variability of this component, correlated with the power-law pivoting
at harder energies, suggest that it is the source of seed photons for
the power-law continuum.
\item A constant reflection component and Fe{\sc xvii} edge can
explain the hard spectral features (in addition to the pivoting
power-law continuum), although a prominent diskline is not required.
However, the apparent lack of long-term coupling between the reflection
amplitude and the narrow iron line
flux, together with the possible variability of the reflection component on
time-scales of weeks suggest that some kind of disk reflection cannot
be ruled out.  The Fe{\sc xvii} edge requires an absorbing column of
$N_{\rm H}>10^{23}$~cm$^{-2}$, but due to the lack of evidence for
such a large absorbing column at lower energies we suggest the edge is
observed in reflection.
\item A constant soft spectral component is also required to explain the spectral
variability of NGC~4051.  This constant soft component likely has a
spectrum similar to that of the variable component, but without any
absorption.  The constant soft component might be related to the
constant hard reflection component if the reflector is ionised, in
which case a model similar to that of \citet{min03b} may help to
explain the unusual spectral variability properties of NGC~4051 in the
low state.  Alternatively, partial covering of the emission by ionised
gas of column density $\sim10^{22}$~cm$^{-2}$ may explain the data.
\end{enumerate}

\subsection*{Acknowledgments}
We thank the anonymous referee for a number of helpful comments which improved this
paper.  This work was supported by grants from the UK Particle Physics and
Astronomy Research Council (PPARC) including PPA/G/S/2000/00085.
IM$^{\rm c}$H also acknowledges the support of a PPARC Senior Research
Fellowship.  The work reported here is based on observations obtained with {\it
XMM-Newton}, an ESA science mission with instruments and contributions
directly funded by ESA Member States and the USA (NASA).

\appendix

\section[]{Scatter in the HS flux-flux relation}
\label{app}
\begin{figure}
\begin{center}
{\epsfxsize 0.9\hsize
 \leavevmode
 \epsffile{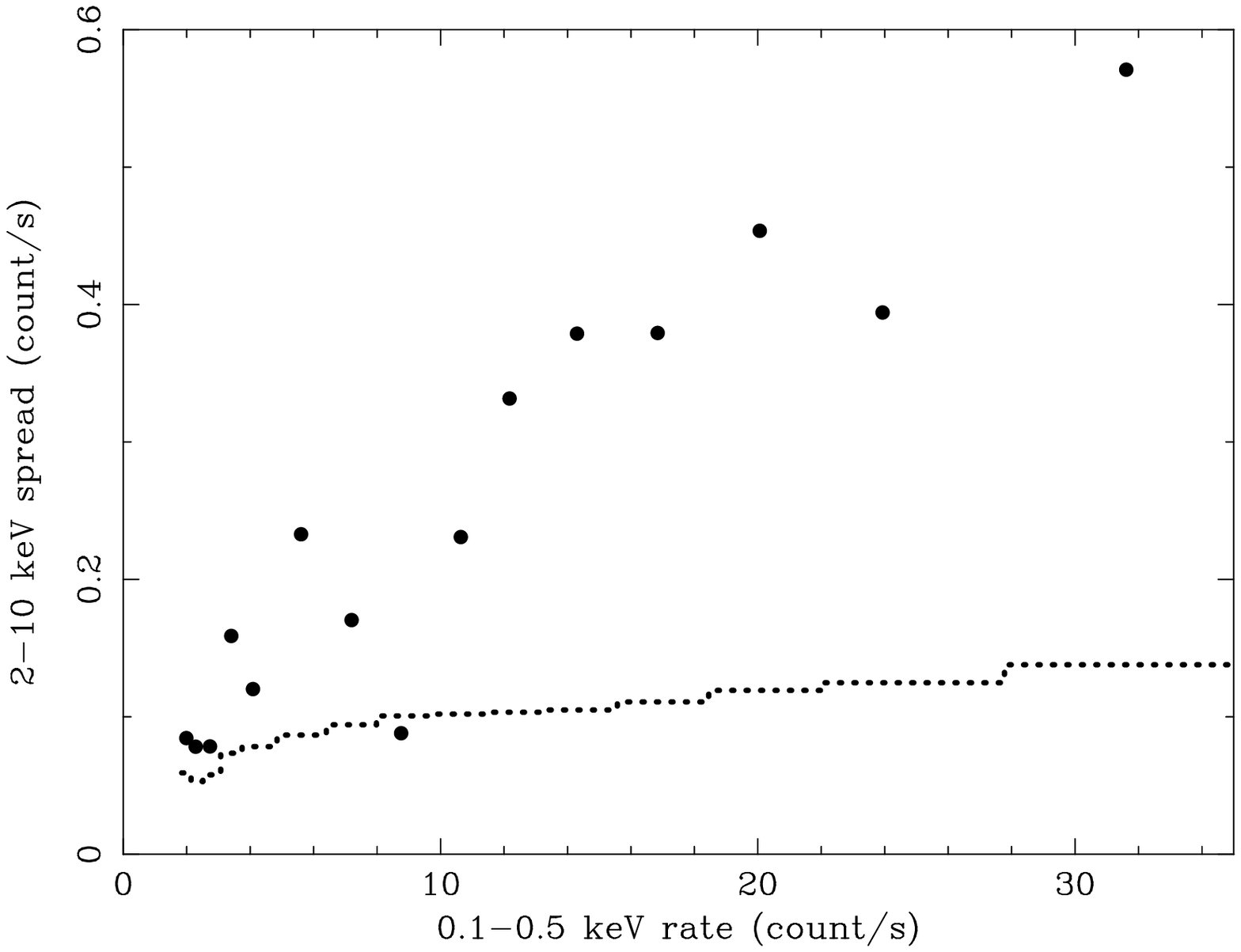}
}\caption{Spread in the HS flux-flux relation (see Section~\ref{fluxflux}
as a function of flux.  The dotted line shows the 90\% confidence
upper limit in spread expected due to photon counting noise effects
only, i.e. with no intrinsic scatter in the unbinned flux-flux relation
(see text for details).} \label{fvfsprd}
\end{center}
\end{figure}
By assuming the best-fitting power-law plus constants model, we can
determine the spread in data points around the binned HS flux-flux relation as a
function of flux, and so confirm that the scatter does increase
towards higher fluxes.  We demonstrate this result in 
Fig.~\ref{fvfsprd}, which shows the rms
spread in the flux bins used to plot the binned HS flux-flux
relation. 
The spread in hard fluxes in a bin, $\sigma_{h}$ is determined using the equation:
 \begin{equation}
\label{eqn:fvfsprd}
\sigma_{h}= \sqrt{\frac{1}{n-1} \sum_{i=1}^{n} (H_{i}-h_{i})^{2}}
\end{equation}
where there are $n$ individual hard-flux measurements ($h_{i}$) in the
bin, and for each individual measurement
$H_{i}$ is the hard-flux predicted from the observed soft flux by the
best-fitting power-law plus
constants model.  Of course, some scatter in the relation can be
caused by photon counting noise in both the hard and soft bands (which
moves the data points on the flux-flux plot in vertical and horizontal
directions respectively).  The effect of this noise on the 
scatter is dependent on the form of the flux-flux relation and so is
not a simple function of flux.  For example, since the fitted
flux-flux relation is a power-law with index$<1$, any
scatter due to noise in the soft band is magnified at low fluxes,
because the flux-flux relation is steeper, so that
a horizontal shift in a data point at low fluxes will have a larger effect
than the same shift at high fluxes.
Therefore, to take account of the effects of photon counting noise, we
used a simple bootstrap-type technique, assigning a hard flux value to each
observed soft flux value using the best-fitting model for the binned HS
flux-flux relation (i.e. assuming no intrinsic scatter)
and then shifting each simulated data point in both the hard flux and soft flux directions
by a zero-mean Gaussian deviate with standard deviation equal to
the corresponding error bars, to produce
a new realisation of the unbinned flux-flux plot.  By simulating 1000
such realisations and measuring the scatter as a function of flux in
each, we could estimate the distribution of the rms spread expected
from photon counting noise only, in the absence of any intrinsic
scatter.  The resulting 90\% confidence upper limit on the expected rms spread
due to noise is shown as a dotted line in Fig.~\ref{fvfsprd}.  Clearly
the observed scatter and the increase in scatter with flux are highly significant.
\end{document}